\documentclass[conference]{IEEEtran}
\usepackage{amsmath,graphicx}
\usepackage[ruled]{algorithm2e}
\usepackage{amssymb}
\usepackage{stfloats} 
\usepackage[english]{babel} 
\usepackage{txfonts}
\usepackage{mathdots}
\usepackage[classicReIm]{kpfonts}
\PassOptionsToPackage{hyphens}{url}

\usepackage[pdftex]{hyperref}

\usepackage{cleveref}
\usepackage{tabularx}
\usepackage{diagbox}
\usepackage{cite}
\usepackage{xcolor}
\usepackage{subfigure}

\hypersetup{
    colorlinks=true,
    allcolors=blue,
    linkcolor=black,
    filecolor=magenta,      
    urlcolor=blue,
    }
\newcommand{\pluseq}{\mathrel{{+}{=}}}

\title{Edge-Parallel Graph Encoder Embedding}

\makeatletter 
\newcommand{\linebreakand}{%
  \end{@IEEEauthorhalign}
  \hfill\mbox{}\par
  \mbox{}\hfill\begin{@IEEEauthorhalign}
}
\makeatother 

\author{\IEEEauthorblockN{Ariel Lubonja}
\IEEEauthorblockA{\textit{Department of Computer Science} \\
\textit{Johns Hopkins University}\\
ariel@cs.jhu.edu}
\and
\IEEEauthorblockN{Cencheng Shen}
\IEEEauthorblockA{\textit{Department of Applied Economics and Statistics} \\
\textit{University of Delaware}\\
shenc@udel.edu}
\linebreakand
\IEEEauthorblockN{Carey Priebe}
\IEEEauthorblockA{\textit{Department of Applied Mathematics and Statistics} \\
\textit{Johns Hopkins University}\\
cep@jhu.edu}
\and
\IEEEauthorblockN{Randal Burns}
\IEEEauthorblockA{\textit{Department of Computer Science} \\
\textit{Johns Hopkins University}\\
randal@cs.jhu.edu}
}

\begin{document}

\maketitle

\begin{abstract}

New algorithms for embedding graphs have reduced the asymptotic complexity of finding low-dimensional representations. One-Hot Graph Encoder Embedding (GEE) uses a single, linear pass over edges and produces an embedding that converges asymptotically to the spectral embedding. The scaling and performance benefits of this approach have been limited by a serial implementation in an interpreted language. We refactor GEE into a parallel program in the Ligra graph engine that maps functions over the edges of the graph and uses lock-free atomic instrutions to prevent data races. On a graph with 1.8B edges, this results in a 500 times speedup over the original implementation and a 17 times speedup over a just-in-time compiled version.

\end{abstract}


\section{Introduction}
\label{sec:intro}

Graph embedding is a powerful technique for exploring the structure of graphs. It is the fundamental step in 
clustering~\cite{clustering-1,  clustering-2}, feature learning~\cite{grover2016node2vec}, and representation learning~\cite{representation-learning} on graphs. 
It is used in the analysis of connectomes~\cite{two-truths}, cybersecurity threat detection~\cite{howie-cybersecurity}, and community detection in social networks~\cite{community-detection}.

Spectral embedding learns a low-dimensional euclidean representation of a graph
\cite{two-truths, spectral-embedding-1} based on a singular-value decomposition (SVD) graph adjacency or graph Laplacian matrix.
Spectral embedding has strong statistical guarantees; the resulting vertex embedding
asymptotically converges to the latent positions under random dot
product graphs \cite{spectral-rdpg} and, thus, is consistent for subsequent inference
tasks such as hypothesis testing and community detection. However, spectral embedding is only as efficient
as the SVD, which is $O(n^3)$ for $n$ vertices and can be solved in $O(n^2 \log n)$ for 
restricted cases or in some approximations \cite{sidestepping-svd}.
Other methods produce good results empirically, but they require parameter tuning or search spaces that are computationally expensive.  Methods based on random walks \cite{grover2016node2vec, DeepWalk} are $O(n)$ but have large constants in the length and number of the walks. 
Graph convolutional neural networks \cite{kipf2016semi} are quite expensive in practice.

Our work focuses on improving the computational performance of the recent one-hot graph encoder embedding (GEE) \cite{GEE}. The algorithm is desirable for its convergence guarantees and because it performs a single pass over the edges.
The base implementation is already an order of magnitude faster than spectral methods, GCN, or node2vec. 
The remaining gap this paper addresses is parallelism and memory efficiency. The current GEE implementation takes nearly an hour on a graph with 1.8B edges. We bring this down to 6.5 seconds. 

We contribute an implementation of GEE that fully utilizes shared-memory hardware and scales to billions of edges. We reformulate the GEE algorithm into an edge-map program in Ligra's programming interface \cite{shun2013ligra}. This {\em GEE-Ligra} implementation avoids data races using lock-free atomic updates, resulting in a 500 times speedup and good scalability. We also provide an optimized version of the
original code using Numba just-in-time compilation. {\em GEE-Ligra} is 5 to 20 times faster than Numba, owing to more efficient
memory usage and multicore parallelism. Our code is publicly available on GitHub: \href{https://github.com/ariellubonja/graph-encoder-embedding}{Numba} and \href{https://github.com/ariellubonja/ligra}{GEE-Ligra}.

\section{Background}
\label{sec:Background}

\noindent {\bf GEE}: For a graph $G(n,s)$ of nodes $n$ connected by edges $s$, the One Hot Graph Encoder Embedding (Algorithm \ref{alg:gee}) builds an embedding matrix $\mathbf{Z}$ based on the edge list $s$ and a vector of class labels $\mathbf{Y}\in \{0,\hdots , K\}^n$, where $K$ is the number of classes.
$\mathbf{Y}$ may represent the labels of a few known node ground truths or it may be derived from unsupervised clustering, such as by running the Leiden community detection algorithm \cite{traag2019leiden}. GEE embeds the $n$ nodes into $k$ dimensions, where $k<<n$.  For brevity, our description does not include the preprocessing steps needed to compute the Laplacian version of the algorithm~\cite{GEE}.

GEE first initiates a projection matrix $\mathbf{W}$ (lines 2-6). Then, in a single pass over the edges, GEE incrementally builds $\mathbf{Z}$ by adding the contribution of each edge to the embedding (lines 7-12).  This contribution is a product of the weight $w$ and the corresponding coefficients in the projection matrix $\mathbf{W}$. The source node contributes to the class of the destination node (line 10) and vice-versa (line 11). This formulation is for weighted directed graphs. Unweighted graphs have unit weights. Undirected graphs are treated as two symmetric directed graphs. 

\vspace{5pt}

\noindent {\bf Ligra}: The {\sf edgeMap}, {\sf vertexMap} interface of Ligra \cite{shun2013ligra} encodes fine-grained, asynchronous parallelism for shared-memory systems. The interface selects a vertex subset, called a {\em frontier}, and then calls a function for every vertex ({\sf vertexMap}) or every outbound edge from the vertex subset ({\sf edgeMap}). This captures almost all modern graph algorithms, including PageRank, Connected Components, and Betweenness Centrality. The frontier subset enables search-style algorithms like breadth-first search (BFS), that trigger computation on neighbors.

\begin{algorithm}[t] 
 \SetKwData{Left}{left}\SetKwData{This}{this}\SetKwData{Up}{up}%
 \SetKwFunction{Union}{Union}\SetKwFunction{GEE}{GEE} \SetKwProg{Fn}{Function}{:}{}
 \SetKwInOut{Input}{input}\SetKwInOut{Output}{output}%
 \Input
  {\begin{minipage}[t]{6cm}
     \strut
     $\mathbf{E} \in \mathbb{R}^{s\times 3}$

     $\mathbf{Y} \in \{0,\hdots , K\}^n$: class labels 
     \strut
   \end{minipage}
  }
 \Output
 {\begin{minipage}[t]{6cm}
 $\mathbf{Z} \in \mathbb{R}^{n\times K}$: node embeddings  
 \strut
   \end{minipage}
 }

 \BlankLine
 \everypar={\nl}
 \Fn{\GEE{$\mathbf{E,Y}$}}{
 
 $\mathbf{W}=zeros(n,K)$     {\color{olive}// Projection matrix}

 \For{$k=1 : K$}%
   {
   {\color{olive}// $k=0$ means class unknown}
   $idx = n \text{ where } Y[n]=k$\\
   $\mathbf{W}(idx,k)=\frac{1}{count(\mathbf{Y}=k)};$
   } 
   \For{$i=1:s$}{
   {\color{olive}// (u-source, v-dest., w-weight)}\\
   $u=\mathbf{E}(i,1);\ v=\mathbf{E}(i,2);\ w=\mathbf{E}(i,3);$\\
   \BlankLine
    $\mathbf{Z}(u,\mathbf{Y}(v))\pluseq \mathbf{W}(v,\mathbf{Y}(v))\cdot w;$\\
    $\mathbf{Z}(v,\mathbf{Y}(u))\pluseq \mathbf{W}(u,\mathbf{Y}(u))\cdot w;$
   }
   }
   \textbf{EndFunction}
 \caption{Semi-Supervised GEE} \label{alg:gee}
 
\end{algorithm}

\section{Methods} \label{GEE Parallelism Approach}

The {\em GEE-Ligra} implementation (Algorithm \ref{alg:ligra}) runs the same algorithm as GEE, i.e.~computes the 
same values on same input. Unlike GEE, which loops over the edges, {\em GEE-Ligra} uses a function map over the edges, which is parallelized and scheduled by the Ligra runtime. the frontier is the entire graph, i.e. all nodes are active. This invokes the {\sf updateEmb} function to update the embedding on all edges in a single step~\cite{shun2013ligra}. This parallelizes GEE's $O(s)$ component. We also parallelize the initialization of the projection matrix, which costs $O(nk)$. For most graphs and choices of $K<50$, $s>nk$. However, $O(nk)$ becomes the dominant component of the runtime when graphs have a high $n$ and a very low average degree.

\begin{algorithm}[t] 
 \SetKwData{Left}{left}\SetKwData{This}{this}\SetKwData{Up}{up}%
 \SetKwFunction{Union}{Union}\SetKwFunction{GEE}{GEE}
 \SetKwFunction{Union}{Union}\SetKwFunction{updateEmb}{updateEmb}
 \SetKwProg{Fn}{Function}{:}{}%
 \SetKwProg{ParFor}{ParallelFor}{ do}{}%
 \SetKwInOut{Input}{input}\SetKwInOut{Output}{output}%
 \Input
  {\begin{minipage}[t]{6cm}%
     \strut
     $\mathbf{E} \in \mathbb{R}^{s\times 3}$, $\mathbf{Y} \in \{0,\hdots , K\}^n$
     \strut
   \end{minipage}%
  }
 \Output{ $\mathbf{Z} \in \mathbb{R}^{n\times K}$, $\mathbf{W} \in \mathbb{R}^{n\times K}$
 }
 \BlankLine
 \BlankLine
 \everypar={\nl}
 \Fn{\GEE{$\mathbf{E,Y}$}}{
 
 $\mathbf{W}=zeros(n,K)$
 
 \ParFor{$k=1 : K$}%
   {
   $idx = n \text{ where } Y[n]=k$\\
   $\mathbf{W}(idx,k)=\frac{1}{count(\mathbf{Y}=k)};$
   }
 \textbf{end}\\
   EdgeMap(\updateEmb, $\mathbf{Z,W,Y}$, frontier=$n$)
   }
   \textbf{EndFunction}
   
   \BlankLine


\BlankLine

\Fn{\updateEmb{$\mathbf{Z, W, Y}, u,v,w$}}{
    writeAdd ($\mathbf{Z}(u,\mathbf{Y}(v)),~ \mathbf{W}(v,\mathbf{Y}(v))\cdot w$ ); \\
    writeAdd ($\mathbf{Z}(v,\mathbf{Y}(u)),~ \mathbf{W}(u,\mathbf{Y}(u))\cdot w$ );
}

\textbf{EndFunction}

 \caption{{\em GEE-Ligra}} \label{alg:ligra}

\end{algorithm}

\begin{figure}[ht]
\vspace{-2pt}
     \centering
     \includegraphics[width=.68\columnwidth]{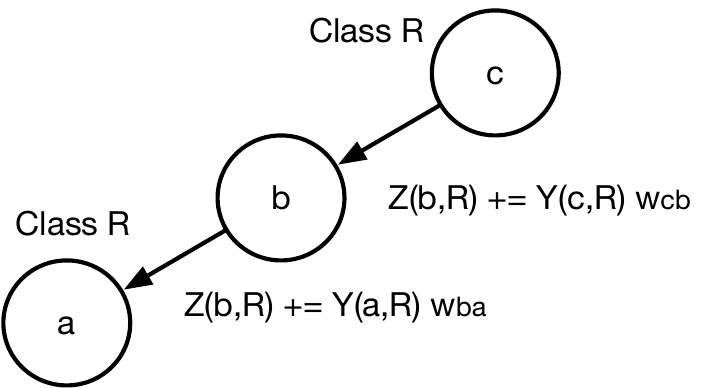}
     \caption{Race condition in which both inbound and outbound edges contribute to the emdedding and create conflicting updates.}
     \label{fig:race}
\vspace{-2pt}

\end{figure}

The {\sf updateEmb} function is mapped across all nodes and access the edge list of each node sequentially.
Because the frontier is the entire graph, Ligra evaluates {\sf updateEmb} using the {\sf edgeMapDense} algorithm \cite{shun2013ligra}. This schedules one worker for the edge list of each node to process all edges sourced from that node sequentially. The read accesses to $\mathbf{W}$ and write accesses to $\mathbf{Z}$ are fine-grained. $\mathbf{Z}(u,:)$ (line 10) and $\mathbf{W}(u,:)$ (line 11) are systematically reused during a {\sf edgeMapDense}, and  will be in the processor cache, however, access to $\mathbf{Z}(v,:)$ and $\mathbf{W}(v,:)$ will likely result in cache misses.

The Ligra {\sf writeAdd()} function uses hardware support to perform a lock-free atomic increment on the embedding field.  This protects the data from the race that occurs when two edges with nodes in the same class update the same entry (Figure \ref{fig:race}). This is caused by GEE propagating class information from source to destination and destination to source.
We expect such conflicts to happen infrequently, because it requires simultaneous scheduling of two separate edges on separate nodes with the same class label. Updates from the $\mathbf{Z}(u,\mathbf{Y}(v_1))$ and $\mathbf{Z}(u,\mathbf{Y}(v_2));~ v_1 \neq v_2, \mathbf{Y}(v_1) == \mathbf{Y}(v_2)$ will not conflict. They are scheduled serially by {\tt edgeMapDense} because they are successive entries in $u$'s edge list.

\begin{figure}[ht]
    \centering
    \includegraphics[width=0.49\textwidth]{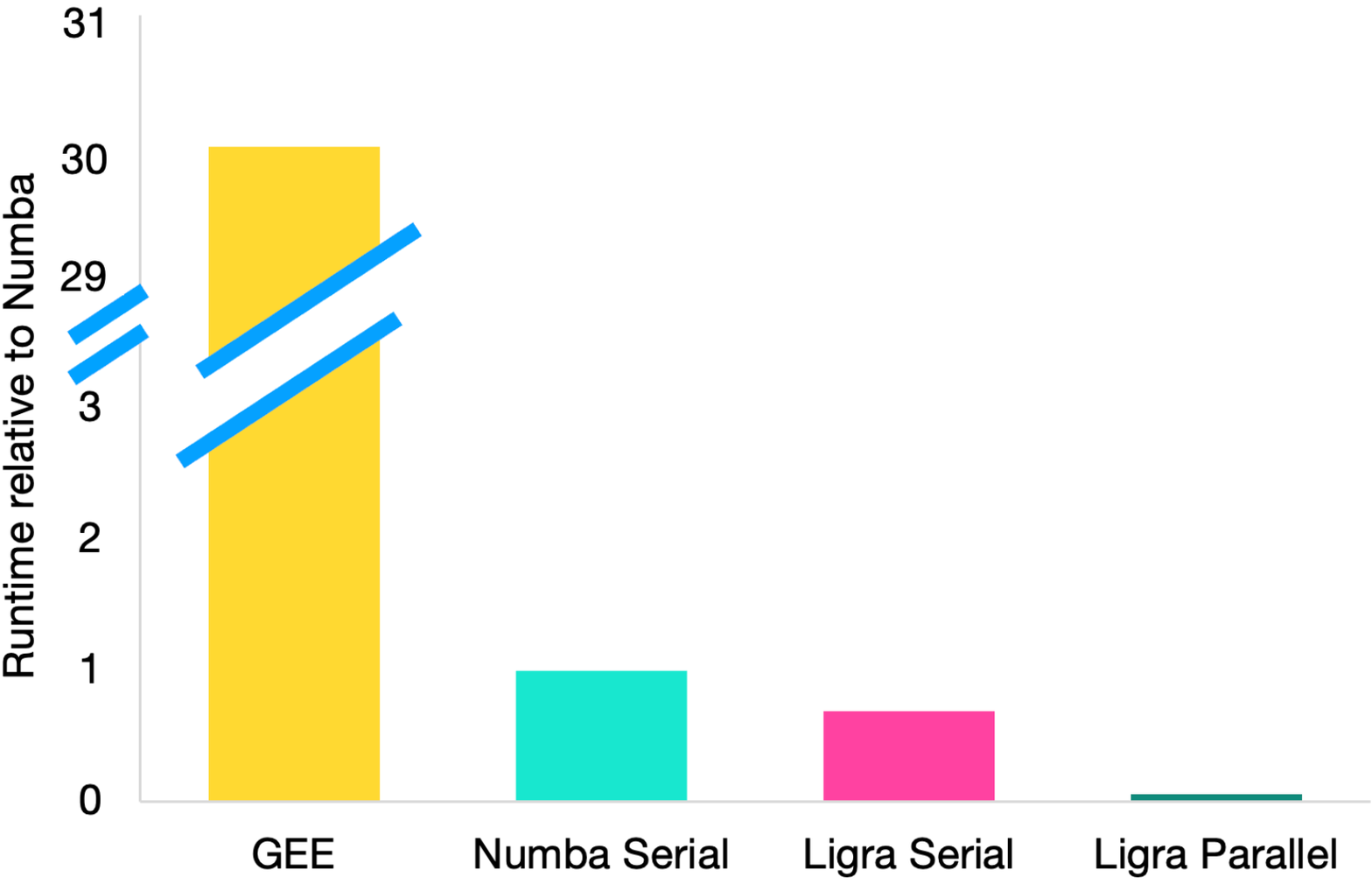}

    \vspace{-5pt}
    
    \caption{Runtimes for Friendster, normalized to Numba Serial. }
    \label{fig:friendster}
\end{figure}

\begin{table*}[t]
\begin{center}
\small
\begin{tabular}{ |c|c|c|c|c|c|c|c|c| } 
 \hline
 \diagbox{Graph ($|n|,|s|)$}{Runtime (sec)} & \shortstack[t]{GEE-\\Python} & \shortstack[t]{Numba\\Serial} & \shortstack[t]{GEE-Ligra\\Serial} & \shortstack[t]{GEE-Ligra\\Parallel} & \shortstack[t]{Speedup\\(v. GEE)} & \shortstack[t]{Speedup\\(v. Numba)} & \shortstack[t]{Speedup\\(v. Ligra Serial)} \\ 

 \hline

 Twitch ($n=168$K, $s=6.8$M) & 12.18 & 0.20 & 0.11 & 0.013 & 936 & 15 & 8.5 \\
 soc-Pokec (1.6M, 30M) & 133.21 & 1.68 & 0.99 & 0.12 & 1100 & 14 & 8.25 \\ 
 soc-LiveJournal (6.4M, 69M) & 301.64 & 4.29 & 2.39 & 0.39 & 773 & 11 & 6.12\\ 
 soc-orkut (3M, 117M) & 499.83 & 4.48 & 2.97 & 0.26 & 1897 & 17 & 11.4\\ 
 orkut-groups (3M, 327M) & 595.29 & 11.43 & 6.06 & 2.36 & 252 & 4.8 & 2.6 \\ 
 Friendster (65M, 1.8B) & 3374.72 & 112.33 & 77.23 & 6.42 & 525 & 17 & 12 \\ 
 \hline

\end{tabular}
\normalsize

\vspace{5pt}
\caption{Runtime (seconds) for graphs of various sizes.  $k=50$ is used for all graphs. Speedup (rows 5-7) show the performance improvement of {\em GEE-Ligra} run in parallel on 24 cores to other implementations.}
\label{table:runtime}
\end{center}
\end{table*}

\section{Findings}\label{Experiments}

We evaluate the parallel implementation of {\em GEE-Ligra} against the reference implementation in Python and our Numba just-in-time compiled implementation. Experiments were performed using a machine with a 24-core, 48-thread Intel Xeon Platinum 8259CL with 192GB main memory. We used Python 3.10.12 and compiled C++ code with the GNU G++ 11.4.0 compiler. Ligra was compiled from \href{https://github.com/jshun/ligra}{source}. 
We experimented with a variety of graphs collected from the SNAP repository~\cite{snapDatasets} that vary between 6.8-327M edges and the Friendster graph \cite{networkrepository} as an example of an Internet-scale dataset with 1.8B edges.
We generated the $\mathbf{Y}$ labels uniformly at random from $[0,K=50]$ for 10\% of nodes, which were also selected uniformly at random.
This practice aligns with GEE's  experimental configuration~\cite{GEE}.

Table \ref{table:runtime} summarizes performance results. Massive performance gains are realized by moving to compiled code. This can be seen through our Numba results which show a 30-50 times speedup. {\em GEE-Ligra} obtains a further 10-15 times performance improvement over Numba. This results from a combination of asynchronous execution in the Ligra graph engine and parallelism. Ligra run in serial improves performance over Numba by less than a factor of 2.

\begin{figure}[ht]
    \centering
    \includegraphics[width=0.49\textwidth]{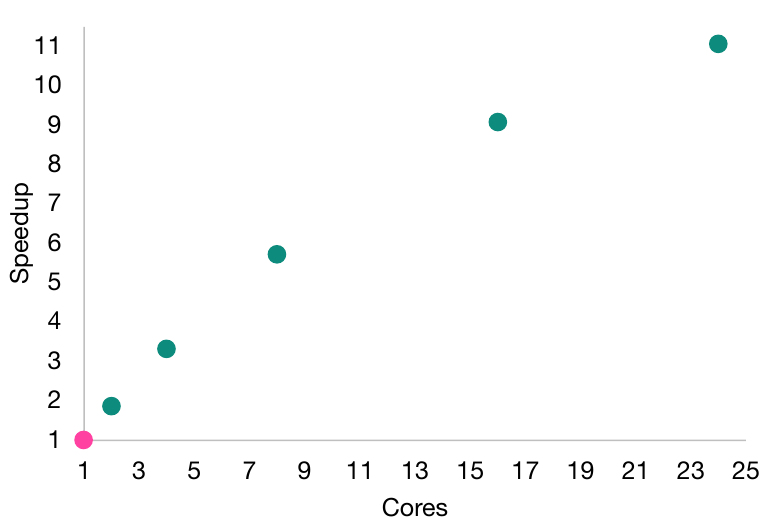}

    \vspace{-5pt}
    
    \caption{Speedup of Ligra on Friendster by nr. cores.}
    \label{fig:strong-scaling}
\end{figure}

The results comparing compiled code to {\em GEE-Ligra} on the largest graph (Figure \ref{fig:friendster}) provide insights into the benefit of using a declarative graph engine. On a single thread on a single core, {\em GEE-Ligra} reduces runtime by 31\%. 
Running {\em GEE-Ligra} in parallel produces a 17 times speedup over Numba and 12 times over {\em GEE-Ligra} serial. 


We study the strong-scaling performance of {\em GEE-Ligra} on the Friendster graph.  The algorithm exhibits good scalability, realizing 11 times speedup over 24 cores (24 threads with hyperthreading disabled). Atomic updates could lead to interference between threads and limit scalability. However, we ran the program with atomics off, performing unsafe updates, and saw no appreciable performance difference. We expect this workload to be memory bound, because there is so little computation per edge. 
GEE-Ligra performs two fused-multiply adds per edge and two memory writes, one of which is likely to miss. These scaling results are consistent with other graph algorithms~\cite{shun2013ligra, wheatman2023cpma} which have been shown to be memory bound. 

\begin{figure}[ht]
    \centering
    \includegraphics[width=0.49\textwidth]{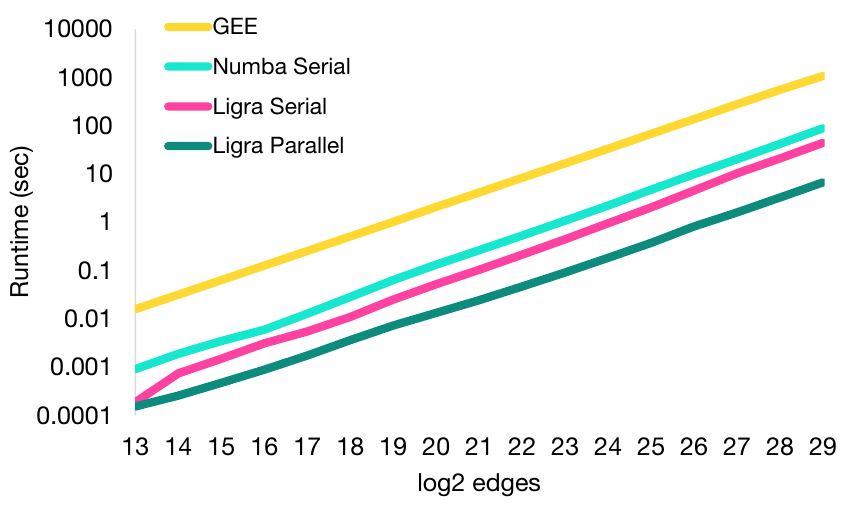}

    \vspace{-5pt}

        \caption{Runtime as the number of edges increase on Erd\H{o}s-R\'enyi graphs.}
    
    \label{fig:weak-scaling}
\end{figure}

We also study a large range of graph sizes to demonstrate that we preserve performance as inputs grow.
We generate Erd\H{o}s-R\'enyi random graphs increasing numbers of edges and run {\em GEE-Ligra} using all 24 cores. Figure \ref{fig:weak-scaling} shows that GEE-Ligra's runtime increases linearly with the number of edges.

\section{Conclusion}
We present an edge-parallel implementation in a shared-memory graph engine of the One Hot Graph Encoding Embedding (GEE) algorithm.
The original algorithm provides an order of magnitude performance improvement over spectral methods, random walks, and graph convolutional networks~\cite{GEE}. Our implementation provides a speedup of 500 times over the base implementation and 17 times speedup over a compiled version of the algorithm. This allows us to embed graphs of 1.8B nodes in 6.5 seconds. Our treatment describes how to parallelize over the edge lists of each node and use atomic instructions to avoid race conditions. 


\bibliographystyle{IEEEbib}
\bibliography{gee}

\end{document}